\documentclass[pre,twocolumn,amsmath,amssymb,superscriptaddress,showpacs,floatfix]{revtex4-2}

\usepackage{graphicx}
\usepackage{amsmath}
\usepackage{amssymb}
\usepackage{hyperref}

\begin{document}

\title{Neural Network Perturbation Theory (NNPT): Learning Residual Corrections from Exact Solutions}

\author{Zhenhao Chen}
\affiliation{Department of Physics, Washington University in St.~Louis, St.~Louis, MO 63130, USA}

\author{Mutian Shen}
\affiliation{Department of Physics, Washington University in St.~Louis, St.~Louis, MO 63130, USA}

\author{Boris Fain}
\affiliation{Freecurve Labs, Berkeley, CA 94710, USA}
\affiliation{Stanford University, Palo Alto CA 94305, USA}

\author{Zohar Nussinov}
\email{corresponding author: zohar@wustl.edu}
\affiliation{Department of Physics, Washington University in St.~Louis, St.~Louis, MO 63130, USA}
\affiliation{Institut f\"{u}r Physik, Technische Universit\"{a}t Chemnitz, 09107 Chemnitz, Germany}
\affiliation{Department of Physics and Quantum Centre of Excellence for Diamond and Emergent Materials (QuCenDiEM), Indian Institute of Technology Madras, Chennai 600036, India}
\affiliation{LPTMC, CNRS-UMR 7600, Sorbonne Universite, 4 Place Jussieu, 75252 Paris cedex 05, France}

\begin{abstract}
Many complex physical systems naturally decompose into an exactly solvable component augmented by a perturbative correction. Rather than directly employing neural networks to analyze complex physical systems, we introduce Neural Network Perturbation Theory (NNPT)---a correction learning approach that predicts residual perturbations after analytically subtracting known mathematically exact solutions. Using the gravitational three-body problem as testbed, we vary Jovian mass from $f=0.05$ to 30 times its physical value while holding network architecture fixed. An equalized-accuracy protocol with 1\% tolerance reveals an unexpected non-monotonic capacity profile: capacity peaks at $f = 5$ in the late integrable regime (3$\times$32 with 2,242 parameters), remains elevated through the transition region ($f \sim 15$--17), then \emph{decreases} in the fully chaotic regime ($f \geq 17$, requiring only 2$\times$32 with 1,186 parameters)---a 47\% reduction from peak. With symplectic integrator energy conservation below $2 \times 10^{-4}$, this counterintuitive phenomenon reflects genuine physical structure rather than numerical artifacts. Sequential correction experiments show negligible refinement ($\|\mathbf{y}_2\| / \|\mathbf{y}_1\| \approx 0.997$), confirming that single-stage networks capture dominant perturbative features without hierarchical decomposition. The capacity transition at $\hat{f}_c = 16.6 \pm 2.8$ aligns with Chirikov's resonance-overlap criterion, suggesting that intermediate-complexity regimes---where perturbations are strong yet dynamics remain quasi-periodic, or where resonances partially overlap---impose maximal capacity requirements, while fully chaotic dynamics undergo \emph{ergodic smoothing}---trajectory-specific fluctuations become irreducible noise, leaving only statistically smooth corrections that require fewer parameters to represent. Our results establish NNPT correction learning as a parameter-efficient strategy and reveal that physical complexity exhibits rich non-monotonic structure that challenges conventional assumptions about chaos and learning difficulty. Crucially, our findings demonstrate that neural network capacity serves as a novel complexity metric that is qualitatively distinct from classical dynamical measures such as Lyapunov exponents: the two decouple at the chaos transition, with dynamical complexity increasing monotonically while representational complexity peaks at intermediate disorder.
\end{abstract}

\maketitle

\section{Introduction}

Neural networks increasingly serve as surrogates for complex physical systems~\cite{breen2020,Liao2022,greydanus2019,cranmer2020}, yet conventional approaches force networks to learn complete solutions from scratch---allocating capacity to both trivial baseline features already known analytically and complex perturbative corrections. In the Sun--Earth--Jupiter system, why should networks relearn the Keplerian ellipse when we seek only Jupiter's perturbation?

We introduce Neural Network Perturbation Theory (NNPT): train networks to predict only the residual $\mathbf{y}(t) = \mathbf{r}_{\text{full}}(t) - \mathbf{r}_{\text{exact}}(t)$ after subtracting known exact solutions. This structure appears ubiquitously and underlies common perturbative approaches. It is, e.g., pervasive in the analysis of weakly nonlinear PDEs, perturbative calculations in quantum mechanics and quantum field theories, mean-field corrections in many-body physics, and numerous other arenas~\cite{BenderOrzag,KC,Holmes,Fernandez,LL,QFT,Fain1}. However, this {\it modus operandi} remains underexplored in machine learning approaches.

To address the fundamental question of how intrinsic dynamical complexity limits representational capacity, we study the gravitational three-body problem---a paradigm for deterministic chaos since Poincar\'{e}'s work~\cite{poincare1890}. Earlier studies of three-body systems employing machine learning~\cite{breen2020,Liao2022} directly tackled the full problem without resorting to the perturbative correction learning approach that we introduce and undertake in the current work. Previous work~\cite{breen2020} trained networks with ${\sim}150$k parameters to learn complete three-body trajectories. In contrast, our correction-learning framework achieves comparable accuracy with ${\sim}1{,}200$ parameters in the integrable regime---a 140-fold reduction by exploiting the known Keplerian baseline. In the Sun--Earth--Jupiter system, varying the Jovian mass reorganizes phase-space structure through resonance overlaps and destruction of invariant tori, central to the Kolmogorov-Arnold-Moser (KAM) theory~\cite{laskar1989,murray1999}. Traditional diagnostics such as Poincar\'{e} sections~\cite{sethna2006} detect chaos only after KAM tori disintegrate. We demonstrate that neural network capacity requirements provide a more sensitive indicator of incipient chaos, detecting resonance-overlap signatures at or before they become visible in traditional geometric diagnostics.

Recent advances illustrate that neural networks may effectively emulate physical systems~\cite{breen2020,Liao2022,sanchez2020,cranmer2020}, with physics-informed architectures that incorporate conservation laws---Hamiltonian neural networks~\cite{greydanus2019}, symplectic networks~\cite{jin2020}, and equivariant networks~\cite{Batzner2022,Kondor2025} which suggest much promise. As rigorously demonstrated long ago, neural networks may, in principle, approximate functions of arbitrarily high complexity~\cite{UAP1,UAP2}. Most studies optimize network architectures for given physical configurations~\cite{raissi2019,karniadakis2021}, conflating model expressivity with inherent dynamical complexity.

We introduce a controlled perturbation experiment using fully connected MLPs with SiLU (Sigmoid Linear Unit)~\cite{elfwing2018} activations across all experiments. The prominent independent variable is that of the simulated Jovian mass (being $f$ times Jupiter's real mass). The dominant Keplerian two-body motion is analytically tractable; Jupiter-induced perturbations constitute genuine three-body effects. We train networks to predict only the residual perturbation
\begin{equation}
\mathbf{y}(t) = \mathbf{r}_E^{(3\text{-body})}(t) - \mathbf{r}_E^{(\text{Kepler})}(t),
\label{eq:residual}
\end{equation}
where $\mathbf{r}_E^{(\text{Kepler})}(t)$ is the exact two-body solution. This decomposition isolates Jupiter's gravitational influence from dominant solar dynamics already captured by Kepler's laws.

To fairly compare learning difficulty across mass factors, we employ an \emph{equalized-accuracy protocol} with 1\% tolerance. A reference 10$\times$128 network (149,122 parameters) defines a mass-independent target validation error $\mathcal{E}_\star$ (median validation MSE across all masses and seeds). For each mass factor $f$, we systematically reduce network capacity until this target accuracy is achieved. Minimal capacity $P(f)$ and training time $\tau(f)$ required to reach $\mathcal{E}_\star$ serve as our primary metrics of learning difficulty.

Both $P(f)$ and $\tau(f)$ vary with $f$ but exhibit a pronounced slope change at $\hat{f}_c = 16.6 \pm 2.8$. Counterintuitively, minimal network capacity \emph{decreases} in the fully chaotic regime: the transition region ($f \sim 15$--17) requires 3$\times$32 architectures (2,242 parameters), while higher masses ($f \geq 17$) achieve target accuracy with smaller 2$\times$32 networks (1,186 parameters). The relative energy drift of our symplectic integrator remains below $2 \times 10^{-4}$ across all $f$, ruling out numerical artifacts. The observed breakpoint aligns with Chirikov's resonance-overlap criterion~\cite{chirikov1979}, which predicts incipient chaos when $f \sim 10$--$16$. Our transition occurs at or before the onset of visible KAM torus distortion in phase-space geometry, suggesting that capacity metrics provide an earlier and more sensitive indicator of incipient chaos than traditional geometric diagnostics.

These results establish several key principles. (i) {\emph{Neural Network Perturbation Theory (NNPT)}} provides a general strategy for parameter-efficient neural surrogates whenever physical systems admit exact decompositions. (ii) \emph{Physical chaos exhibits non-monotonic complexity structure}: the transitional regime where resonances partially overlap imposes the highest capacity requirements, while fully chaotic dynamics paradoxically yield simpler residual corrections that require fewer parameters. (iii) \emph{Neural network capacity and classical dynamical complexity decouple at the chaos transition}: while Lyapunov exponents and Kolmogorov--Sinai entropy increase monotonically with $f$, representational capacity peaks at intermediate disorder---revealing that these two notions of complexity measure fundamentally different properties of the system. Understanding this non-monotonic relationship is essential for deploying neural surrogates in computational physics, where purely data-driven black-box approaches may face unexpected scaling challenges precisely in intermediate-complexity regimes.

\begin{figure*}[tb]
\centering
\includegraphics[width=0.95\textwidth]{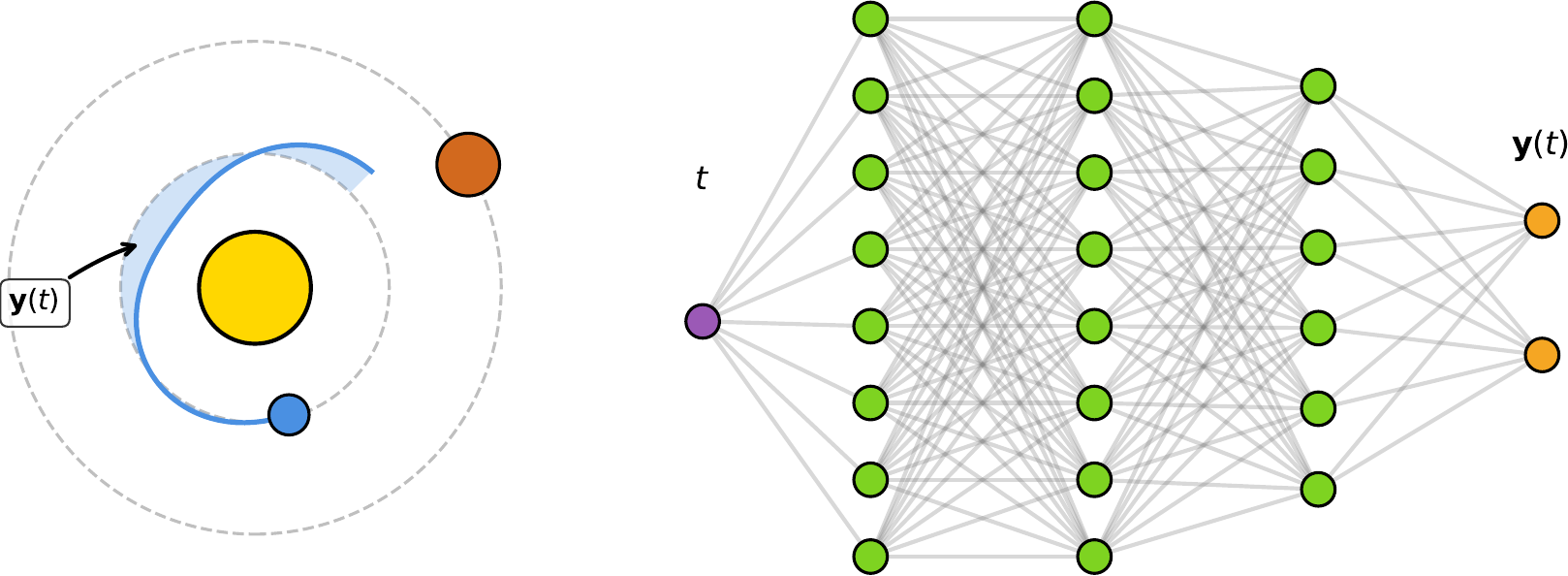}
\caption{Neural Network Perturbation Theory (NNPT) framework for the three-body problem. (Left) Schematic of the Sun--Earth--Jupiter system showing Earth's perturbed orbit (solid blue curve) deviating from the exact Keplerian circular orbit (dashed gray). The shaded region $\mathbf{y}(t)$ represents the residual perturbation---the difference between the three-body trajectory and the analytical two-body solution. (Right) Multilayer perceptron architecture (simplified visualization: 3 hidden layers with 6--8 neurons per layer) that learns only the residual correction $\mathbf{y}(t)$ from normalized time input $t$. The network receives time $t \in [0,1]$ and outputs the two-dimensional perturbation $\mathbf{y}(t)$, focusing representational capacity exclusively on genuine three-body dynamics rather than redundantly approximating the known Keplerian baseline.}
\label{fig:nnpt_schematic}
\end{figure*}

\section{Methods}
\label{sec:methods}

\subsection{Physical System and Parametrization}

We study the planar circular restricted three-body problem consisting of the Sun (mass $M_\odot$), Earth (mass $M_\oplus = 3.003 \times 10^{-6} M_\odot$), and Jupiter (variable mass $m_J$). We work in heliocentric coordinates with astronomical units: distances in AU, time in years, masses in solar masses, with gravitational constant $G = 4\pi^2$~AU$^3$ M$_\odot^{-1}$ yr$^{-2}$.

The system is initialized with Earth and Jupiter in circular coplanar orbits with semi-major axes $a_E = 1.0$~AU and $a_J = 5.2$~AU. Earth's eccentricity is $e_E = 0.0167$. The Jovian mass was parametrized as
\begin{equation}
m_J = f \cdot m_J^{(\text{real})}, \quad m_J^{(\text{real})} = 9.548 \times 10^{-4} M_\odot.
\end{equation}
The mass factor $f$ is varied over $[0.05, 30]$, sampling 32 points with particularly dense coverage in the transition region ($f = 13$--20) where capacity requirements are expected to change rapidly based on Chirikov's resonance-overlap criterion.

At $f = 1$, the system approximates the real solar system. For $f \ll 1$, Jupiter acts as a weak perturbation and Earth's motion is nearly integrable. For $f \gg 1$, strong gravitational interactions lead to chaotic dynamics. Chirikov's resonance-overlap criterion~\cite{chirikov1979} predicts global chaos when $m_J \gtrsim 10m_J^{(\text{real})}$ (i.e., $f \sim 10$--$16$).

\subsection{Numerical Integration}

Trajectories are integrated using the velocity Verlet algorithm, a second-order symplectic method~\cite{hairer2006}. For a system with positions $\mathbf{q}$ and momenta $\mathbf{p}$, the scheme updates
\begin{align}
\mathbf{p}_{n+1/2} &= \mathbf{p}_n + \frac{\Delta t}{2} \mathbf{F}(\mathbf{q}_n), \\
\mathbf{q}_{n+1} &= \mathbf{q}_n + \Delta t \, \mathbf{p}_{n+1/2}/m, \\
\mathbf{p}_{n+1} &= \mathbf{p}_{n+1/2} + \frac{\Delta t}{2} \mathbf{F}(\mathbf{q}_{n+1}),
\end{align}
where $\mathbf{F}$ is the gravitational force. We integrate over $T = 20$~yr with base time step $\Delta t = 8 \times 10^{-3}$~yr, yielding $N = 2500$ time points per trajectory.

Energy conservation is monitored via the dimensionless relative drift
\begin{equation}
\delta E = \frac{|E(T) - E(0)|}{|E(0)|},
\end{equation}
where $E = K + U$ is total energy. If $\delta E$ exceeds $10^{-6}$, the segment is reintegrated with $\Delta t \to \Delta t/2$. In practice, adaptive refinement is rarely triggered ($<2\%$ of trajectories), and final drift remains below $2 \times 10^{-4}$ for all $f$, and below $2 \times 10^{-7}$ for $f \geq 8$ (Fig.~\ref{fig:energy}). This exceptional energy conservation ensures observed transitions reflect genuine physical complexity rather than numerical artifacts.

The Keplerian baseline $\mathbf{r}_E^{(\text{Kepler})}(t)$ is computed analytically by solving Kepler's equation with orbital elements matching initial conditions~\cite{murray1999}.

\subsection{Neural Network Perturbation Theory (NNPT)}

We now explicitly turn to the core idea underlying our analysis. A standard approach trains networks to learn the complete three-body trajectory $\mathbf{r}_E^{(3\text{-body})}(t)$ directly from time input, requiring networks to represent both (1) the dominant Keplerian ellipse---already exactly solvable---and (2) Jupiter's subtle perturbative corrections. Our correction-learning NNPT framework exploits the natural decomposition of Eq.~(\ref{eq:residual}) where $\mathbf{r}_E^{(\text{Kepler})}(t)$ is computed analytically and only the residual $\mathbf{y}(t)$ is learned. By training networks to predict only $\mathbf{y}(t)$ rather than the full trajectory, we isolate genuinely complex three-body dynamics from the tractable Keplerian baseline. The conceptual foundation is directly analogous to standard perturbation theory, with the network encapsulating all subsequent expansion terms. The architecture and decomposition are illustrated in Fig.~\ref{fig:nnpt_schematic}.

The perturbative correction $\mathbf{y}(t)$ has a substantially lower amplitude than the complete trajectory $\mathbf{r}_E^{(3\text{-body})}(t)$ (typically $\|\mathbf{y}\| \ll \|\mathbf{r}_E\|$), thus dramatically reducing the dynamic range that the networks must represent. More fundamentally, network capacity is allocated exclusively to physically meaningful perturbations---genuine three-body effects---rather than redundantly approximating the well-understood Keplerian baseline. This yields over 140-fold parameter reduction compared to full-trajectory approaches~\cite{breen2020}. Before further discussing details of the three-body gravitational problem at hand, we remark that the simple concept that we employ might generalize to multiple other physical systems admitting a mathematically rigorous and exact solution that is then augmented by perturbative corrections. This method may, in principle, be applied to various problems in, e.g.,

\begin{itemize}
\item \textbf{Quantum mechanics}: Learn corrections to Hartree-Fock, density functional theory, or exactly solvable model Hamiltonians to capture correlation effects.
\item \textbf{Nonlinear PDEs}: Learn corrections to linearized or weakly nonlinear solutions to refine analytical approximations.
\item \textbf{Many-body physics}: Learn correlation corrections beyond mean-field approximations to capture fluctuations and collective modes.
\item \textbf{Molecular dynamics}: Learn atomistic corrections to coarse-grained continuum solutions; learn residual QM-MM Hamiltonian correspondence after physics-inspired terms (VdW, electrostatics); learn PIMD corrections to classical trajectory~\cite{Fain1}.
\item \textbf{Plasma physics}: Learn kinetic corrections to magnetohydrodynamic approximations to capture wave-particle interactions.
\end{itemize}

The key conceptual shift in our approach which might, hopefully, be useful to problems in the above and other domains, is that instead of inquiring whether ``can a neural network learn this complex system?'', we ask ``given that we already know the dominant behavior analytically, can a neural network learn only the interesting residual?'' Such a reframing may recast challenging scientific problems for which the insights found by currently employed neural networks can be somewhat opaque and limited into more manageable forms.

\subsection{Training Data}

For each $f$, we generate training and validation datasets by integrating the three-body system over $T = 20$~yr. The network does \emph{not} learn to predict the full Earth position $\mathbf{r}_E^{(3\text{-body})}(t)$. Instead, we compute the exact Keplerian baseline $\mathbf{r}_E^{(\text{Kepler})}(t)$ analytically for each time point, then form the correction signal $\mathbf{y}(t) = \mathbf{r}_E^{(3\text{-body})}(t) - \mathbf{r}_E^{(\text{Kepler})}(t)$ as defined in Eq.~\eqref{eq:residual}. Network input is normalized time $\tilde{t} = t/T \in [0,1]$, and output is this two-dimensional residual $\mathbf{y}(t)$---encoding only Jupiter's gravitational influence. We use an 80\%/20\% train/validation split, yielding 2000 training and 500 validation samples per trajectory.

\subsection{Neural Network Architecture}

We employ a fully connected feedforward MLP with fixed architecture across all baseline experiments:
\begin{itemize}
\item Input: 1 unit (normalized time $\tilde{t}$)
\item Hidden: The largest reference system that we employed was that of 10 layers $\times$ 128 units per layer. Experiments on multiple other architectures were performed (see Table~\ref{tab:architecture_grid}).
\item Output: 2 units (Cartesian components of $\mathbf{y}$)
\item Activation: SiLU
\item Total parameters: 149,122
\end{itemize}

The SiLU activation $\sigma(x) = x/(1 + e^{-x})$ is chosen for its smooth, non-monotonic profile. No batch normalization or dropout is used, ensuring expressivity is determined solely by depth and width.

\textbf{Architecture Notation:} Throughout this paper, we use depth$\times$width to denote network architectures, where ``depth'' refers to the number of hidden layers and ``width'' to units per hidden layer. For example, 2$\times$32 denotes a network with 2 hidden layers of 32 units each (plus input and output layers), totaling 1,186 trainable parameters for our input dimension 1 and output dimension 2 configuration.

\subsection{Architecture Grid and Parameter Calculation}

To systematically explore capacity requirements, we evaluate networks from the following architecture grid (Table~\ref{tab:architecture_grid}). The grid is designed with three considerations: (i) logarithmic spacing in the low-capacity regime (1k--3k parameters) to efficiently explore the integrable limit; (ii) dense linear sampling in the transition region (3k--16k parameters) where capacity requirements are expected to change sharply based on Chirikov's resonance-overlap criterion; (iii) sparse sampling in the high-capacity regime ($>$16k parameters) where all architectures successfully meet the target accuracy.

The parameter count for a network with $L$ hidden layers, $W$ units per layer, input dimension $d_{\text{in}}=1$, and output dimension $d_{\text{out}}=2$ is:
\begin{eqnarray}
P = (d_{\text{in}} \cdot W + W) + (L-1)(W^2 + W) \nonumber \\
+ (W \cdot d_{\text{out}} + d_{\text{out}}).
\label{eq:param_count}
\end{eqnarray}

\textbf{Critical Implementation Detail:} Although Table~\ref{tab:architecture_grid} lists architectures grouped by depth for clarity, the capacity shrinking procedure tests architectures \emph{strictly in ascending order of total parameter count}, not by architecture name. This parameter-count ordering is essential because architectures with different depth-width combinations can have similar or inverted parameter counts (e.g., 3$\times$32 has fewer parameters than 2$\times$48, and 4$\times$48 has fewer than 3$\times$64). Testing in parameter order prevents spurious capacity overestimation and ensures our reported minimal capacities represent true lower bounds for the equalized-accuracy protocol.

\begin{table}[h]
\centering
\caption{Architecture grid explored in capacity shrinking experiments. Architectures are grouped by depth for presentation clarity, but the shrinking procedure tests them in strict ascending order of parameter count. Parameter counts follow Eq.~\eqref{eq:param_count}.}
\label{tab:architecture_grid}
\begin{tabular}{lcr}
\hline\hline
Architecture & Layers $\times$ Width & Parameters \\
\hline
2$\times$32  & 2 $\times$ 32  & 1,186 \\
2$\times$40  & 2 $\times$ 40  & 1,602 \\
3$\times$32  & 3 $\times$ 32  & 2,242 \\
2$\times$48  & 2 $\times$ 48  & 2,546 \\
3$\times$40  & 3 $\times$ 40  & 3,442 \\
2$\times$56  & 2 $\times$ 56  & 3,490 \\
2$\times$64  & 2 $\times$ 64  & 4,434 \\
3$\times$48  & 3 $\times$ 48  & 5,138 \\
3$\times$56  & 3 $\times$ 56  & 6,850 \\
4$\times$48  & 4 $\times$ 48  & 7,776 \\
3$\times$64  & 3 $\times$ 64  & 8,578 \\
4$\times$56  & 4 $\times$ 56  & 10,248 \\
4$\times$64  & 4 $\times$ 64  & 12,738 \\
5$\times$64  & 5 $\times$ 64  & 16,898 \\
5$\times$80  & 5 $\times$ 80  & 25,922 \\
6$\times$80  & 6 $\times$ 80  & 32,482 \\
6$\times$96  & 6 $\times$ 96  & 46,946 \\
7$\times$96  & 7 $\times$ 96  & 56,482 \\
8$\times$96  & 8 $\times$ 96  & 66,018 \\
10$\times$128 & 10 $\times$ 128 & 149,122 \\
\hline\hline
\end{tabular}
\end{table}

\subsection{Training Protocol}

Networks are trained using the Adam optimizer~\cite{kingma2015} with initial learning rate $\eta_0 = 10^{-3}$ and exponential decay $\eta(e) = \eta_0 \exp(-e/1000)$. The batch size is 64. Training proceeded for up to 5000 epochs with early stopping (patience = 500 epochs). The loss function is the mean squared error,
\begin{equation}
\mathcal{L} = \frac{1}{N_{\text{batch}}} \sum_{i} \|\mathbf{y}_i - \hat{\mathbf{y}}_i\|^2.
\end{equation}
All experiments were repeated over three initial seeds (2025, 2026, 2027) to ensure statistical robustness.

\subsection{Equalized-Accuracy Protocol}
\label{sec:equalized_accuracy}

To isolate the effect of physical complexity, we introduce an equalized-accuracy protocol. First, we train the reference 10$\times$128 network on all mass factors and compute the median validation MSE across all $(f, \text{seed})$ pairs:
\begin{equation}
\mathcal{E}_\star = \text{median}\{\text{ValMSE}(f, s) : f \in \mathcal{F}, s \in \mathcal{S}\}.
\label{eq:target_mse}
\end{equation}
We then define acceptance threshold $\mathcal{E}_{\text{thr}} = \mathcal{E}_\star(1 + \varepsilon)$ with $\varepsilon = 0.01$, imposing a strict 1\% tolerance to ensure meaningful capacity comparisons across mass factors.

For each $(f, s)$ pair, we systematically reduce network capacity by testing architectures from Table~\ref{tab:architecture_grid} in ascending order of parameter count until an architecture satisfies validation MSE $\leq \mathcal{E}_{\text{thr}}$ within 5000 training epochs. For each tested architecture, we record:
\begin{itemize}
\item \textbf{First-hit epoch} $\tau(f, s)$: the epoch at which validation MSE first drops below $\mathcal{E}_{\text{thr}}$.
\item \textbf{Picked capacity} $P(f, s)$: the minimal parameter count achieving $\mathcal{E}_{\text{thr}}$ within 5000 epochs.
\end{itemize}
If no architecture meets the threshold, we report the best-performing configuration.

\subsection{Change-Point Detection}

To identify the transition mass $\hat{f}_c$, we fit a piecewise linear model with a single breakpoint to the sequence of mean picked parameters $\bar{P}(f)$:
\begin{equation}
\bar{P}(f) = \begin{cases}
\alpha_1 + \beta_1 f, & f < f_c, \\
\alpha_2 + \beta_2 f, & f \geq f_c.
\end{cases}
\end{equation}
The breakpoint is determined by minimizing the Bayesian Information Criterion (BIC). Confidence intervals are obtained via bootstrap resampling (200 iterations), computing the 16th and 84th percentiles of the breakpoint distribution (68\% CI).

\subsection{Sequential Perturbative Corrections}

Motivated by higher-order perturbation theory, we explore whether capacity barriers can be circumvented through sequential refinement. We implement a two-stage approach testing nine representative mass factors spanning integrable ($f = 5, 10$), transitional ($f = 14, 15, 16, 17, 18, 20$), and chaotic ($f = 30$) regimes:
\begin{align}
\text{Stage 1:} \quad & \mathbf{y}_1(t) = \mathbf{r}_E^{(3\text{-body})}(t) - \mathbf{r}_E^{(\text{Kepler})}(t), \nonumber \\
& \text{NN}_1 \text{ (3$\times$64) trained to predict } \mathbf{y}_1(t); \\
\text{Stage 2:} \quad & \mathbf{y}_2(t) = \mathbf{y}_1(t) - \text{NN}_1(t), \nonumber \\
& \text{NN}_2 \text{ (2$\times$32) trained to predict } \mathbf{y}_2(t).
\end{align}
We quantify refinement efficiency via the ratio $\|\mathbf{y}_2\| / \|\mathbf{y}_1\|$. Values near unity indicate negligible improvement from sequential correction.

\section{Results}
\label{sec:results}

\subsection{Baseline Performance and Numerical Stability}

\begin{figure}[tb]
\centering
\includegraphics[width=0.48\textwidth]{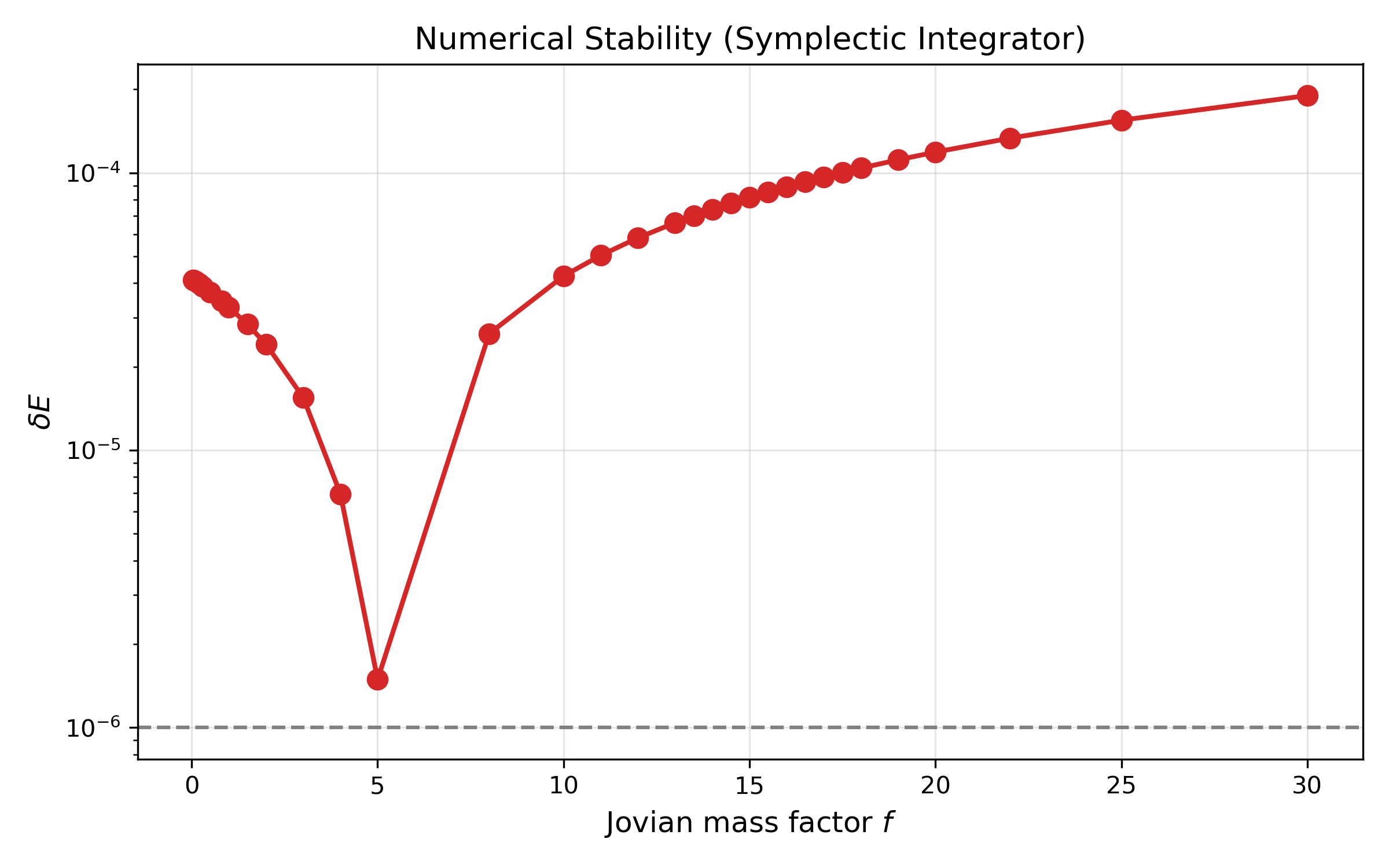}
\caption{Relative energy drift $|\Delta E|/|E_0|$ vs.\ Jovian mass factor $f$ for the symplectic velocity-Verlet integrator. The drift decreases from $\sim\!4\times10^{-5}$ at small $f$ to a minimum near $f=5$---indicative of a stable resonance configuration---then increases monotonically to $\sim\!2\times10^{-4}$ at $f=30$. All values remain well below the adaptive-refinement threshold $10^{-6}$ (dashed line) for $f\geq 8$, confirming that observed capacity transitions reflect genuine physical complexity rather than numerical artifacts.}
\label{fig:energy}
\end{figure}

Figure~\ref{fig:energy} shows the relative energy drift $\delta E$ as a function of mass factor $f$. The drift exhibits a non-monotonic profile: it decreases from $\delta E \approx 4 \times 10^{-5}$ at $f = 0.05$ to a minimum near $f = 5$---consistent with a particularly stable orbital resonance configuration at this mass factor---before increasing monotonically to $\delta E \approx 2 \times 10^{-4}$ at $f = 30$. All values remain well below $10^{-6}$ for $f \geq 8$, and below $2 \times 10^{-4}$ throughout. This confirms excellent energy conservation throughout the parameter range, ruling out numerical artifacts as the source of observed transitions in learning difficulty.

\begin{figure}[tb]
\centering
\includegraphics[width=0.48\textwidth]{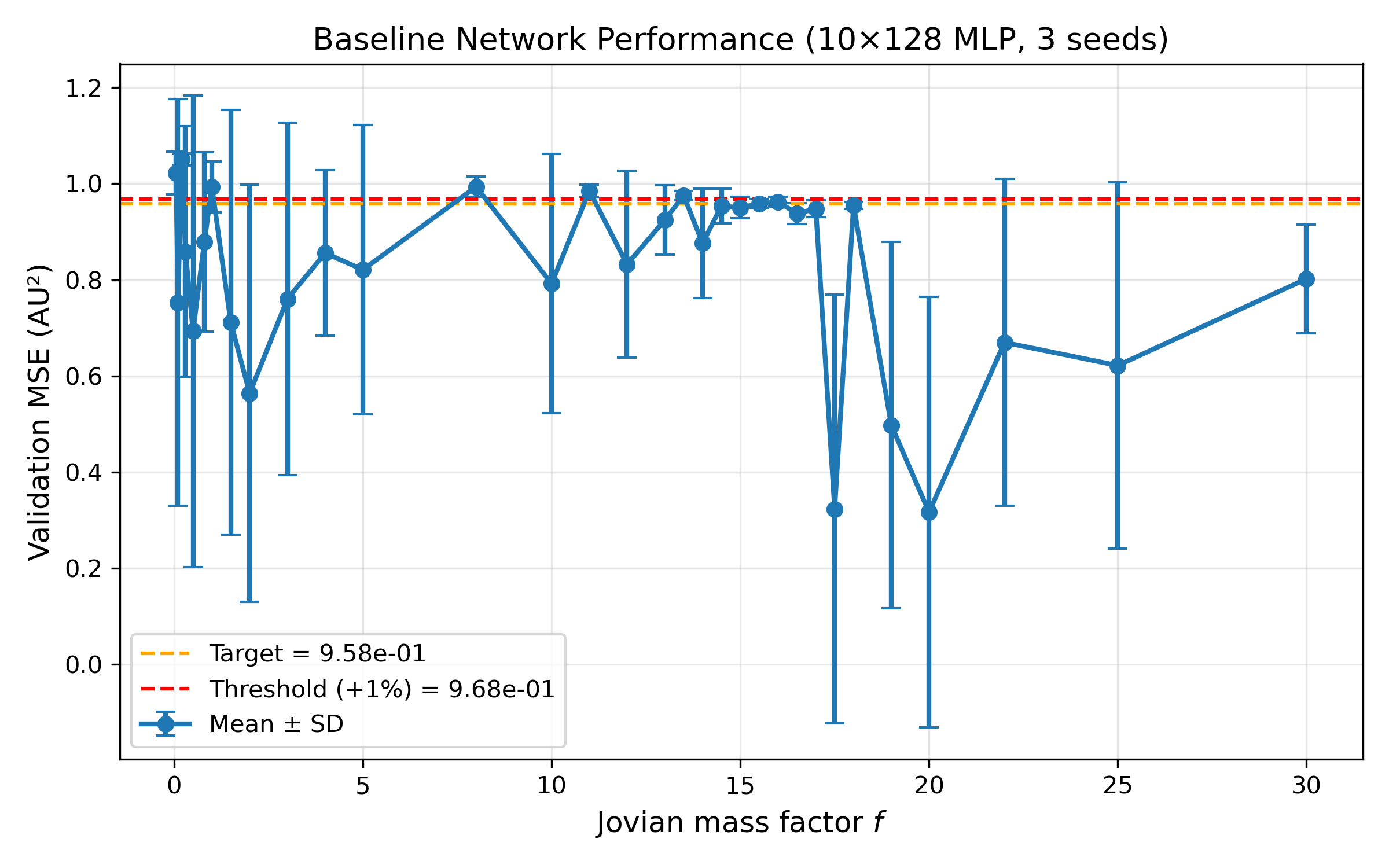}
\caption{Baseline network performance using fixed 10$\times$128 architecture (149,122 parameters) learning correction signal $\mathbf{y}(t)$ across mass factors. Faint lines show individual seeds (2025, 2026, 2027); bold line with error bars indicates mean $\pm$ SD ($n=3$). Horizontal lines mark target MSE $\mathcal{E}_\star = 0.958$ (dashed orange) and acceptance threshold $\mathcal{E}_{\text{thr}} = 0.968$ (dashed red).}
\label{fig:baseline_mse}
\end{figure}

Figure~\ref{fig:baseline_mse} presents the validation MSE of the reference 10$\times$128 network as a function of $f$. The baseline network achieves target accuracy $\mathcal{E}_\star = 0.958$ (median across all masses and seeds) with 1\% acceptance threshold $\mathcal{E}_{\text{thr}} = 0.968$. The MSE profile is relatively flat across most mass factors, with slight elevation in the low-mass regime ($f < 1$) where the residual signal has very small amplitude. A small number of $(f, \text{seed})$ pairs yield anomalously low MSE values ($\ll 0.1$), likely due to fortuitous optimization trajectories; these do not affect $\mathcal{E}_\star$, which is defined as the median and is robust to such outliers.

\subsection{Reverse Capacity Transition}

\begin{figure*}[tb]
\centering
\includegraphics[width=0.95\textwidth]{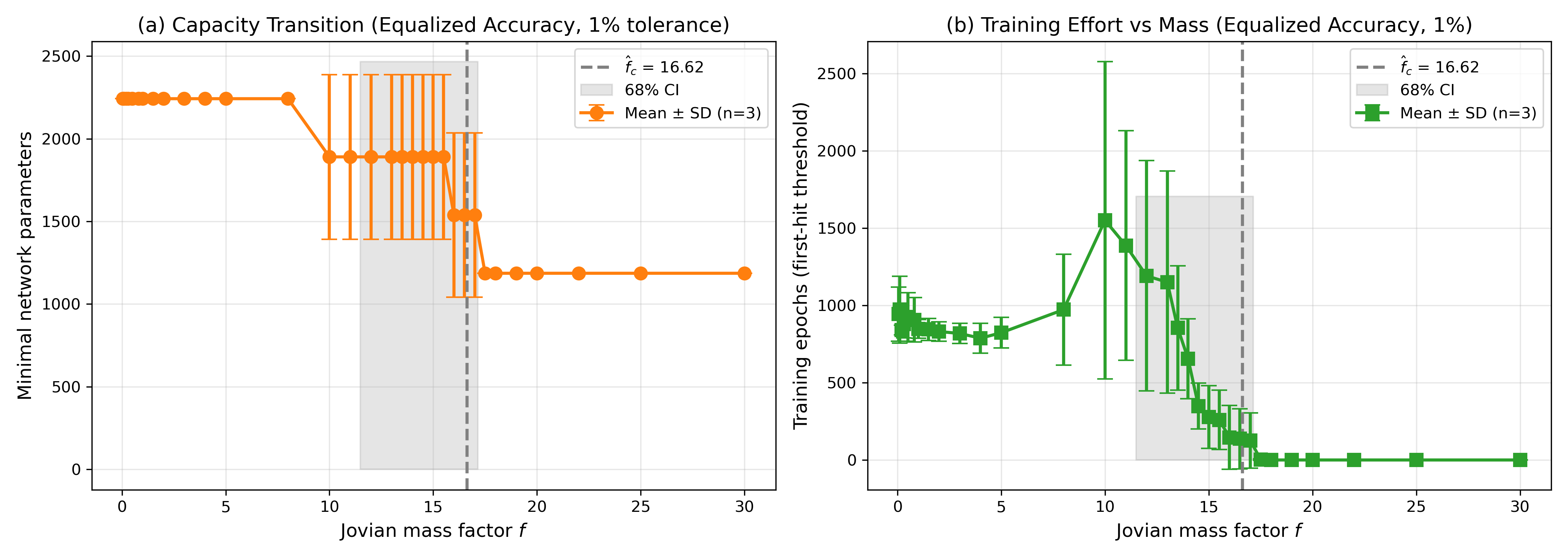}
\caption{Reverse capacity transition at chaos onset under equalized-accuracy protocol with 1\% tolerance. (a)~Minimal network capacity vs.\ Jovian mass factor (orange circles: mean $\pm$ SD over three random seeds; vertical dashed line: breakpoint $\hat{f}_c = 16.6$; gray band: 68\% bootstrap CI [11.5, 17.1]). Capacity peaks at $f = 5$ in the integrable regime (3$\times$32, 2,242 parameters), remains elevated through the transition region ($f \sim 10$--17), and \emph{decreases} in the fully chaotic regime ($f \geq 17$, requiring only 2$\times$32 with 1,186 parameters). For $f \leq 8$ all three seeds consistently select 3$\times$32. The wide CI reflects seed-dependent variability in the transition region---a signature of chaos onset. (b)~Training effort (first-hit epoch) vs.\ mass factor (green squares: mean $\pm$ SD). Training time peaks in the transition region, mirroring the capacity profile.}
\label{fig:capacity_combined}
\end{figure*}

Figure~\ref{fig:capacity_combined} presents our central finding: minimal network capacity exhibits a non-monotonic (reverse) transition, decreasing in the chaotic regime contrary to conventional intuition.

The transition is quantified via piecewise linear regression with breakpoint $\hat{f}_c = 16.6$ (BIC-minimized). Bootstrap resampling yields 68\% CI $[11.5, 17.1]$. Table~\ref{tab:picked_architectures} quantifies the reverse transition in detail. For $f \leq 8$, all three seeds consistently select 3$\times$32 (mean 2,242, std~=~0), reflecting a strictly quasi-periodic regime. At $f = 10$, seed variability begins (mean 1,890~$\pm$~610). By $f = 16$--17, the dominant architecture shifts to 2$\times$32, and the chaotic regime ($f \geq 18$) returns to full 3/3 convergence on 2$\times$32 (1,186 parameters)---a 47\% reduction from peak.

\begin{table*}[tb]
\centering
\caption{Representative minimal network architectures picked by equalized-accuracy protocol with 1\% tolerance. Mean $\pm$ SD over three random seeds. The reverse transition is evident: global capacity peaks at $f=5$, remains elevated through $f\sim14$--17, and decreases sharply in the fully chaotic regime.}
\label{tab:picked_architectures}
\begin{tabular}{cllcr}
\hline\hline
Mass & Dominant & Seed & Parameters & Relative \\
Factor $f$ & Architecture & Convergence & (Mean $\pm$ SD) & to Baseline \\
\hline
\multicolumn{5}{c}{\textit{Integrable Regime}} \\
\hline
5.0  & 3$\times$32 & 3/3 & 2,242 $\pm$ 0 & 1.00$\times$ \\
10.0 & 3$\times$32 (2/3) & Mixed & 1,890 $\pm$ 610 & 0.84$\times$ \\
     & 2$\times$32 (1/3) &  &  &  \\
\hline
\multicolumn{5}{c}{\textit{Transition Region}} \\
\hline
14.0 & 3$\times$32 (2/3) & Mixed & 1,890 $\pm$ 610 & 0.84$\times$ \\
     & 2$\times$32 (1/3) &  &  &  \\
15.0 & 3$\times$32 (2/3) & Mixed & 1,890 $\pm$ 610 & 0.84$\times$ \\
     & 2$\times$32 (1/3) &  &  &  \\
16.0 & 2$\times$32 (2/3) & Mixed & 1,538 $\pm$ 610 & 0.69$\times$ \\
     & 3$\times$32 (1/3) &  &  &  \\
17.0 & 2$\times$32 (2/3) & Mixed & 1,538 $\pm$ 610 & 0.69$\times$ \\
     & 3$\times$32 (1/3) &  &  &  \\
\hline
\multicolumn{5}{c}{\textit{Chaotic Regime}} \\
\hline
18.0 & 2$\times$32 & 3/3 & 1,186 $\pm$ 0 & 0.53$\times$ \\
20.0 & 2$\times$32 & 3/3 & 1,186 $\pm$ 0 & 0.53$\times$ \\
30.0 & 2$\times$32 & 3/3 & 1,186 $\pm$ 0 & 0.53$\times$ \\
\hline\hline
\multicolumn{5}{l}{\small Relative capacity normalized to global peak (3$\times$32, 2,242 parameters, at $f=5$).} \\
\end{tabular}
\end{table*}

\subsection{Sequential Correction Analysis}

\begin{figure*}[tb]
\centering
\includegraphics[width=0.95\textwidth]{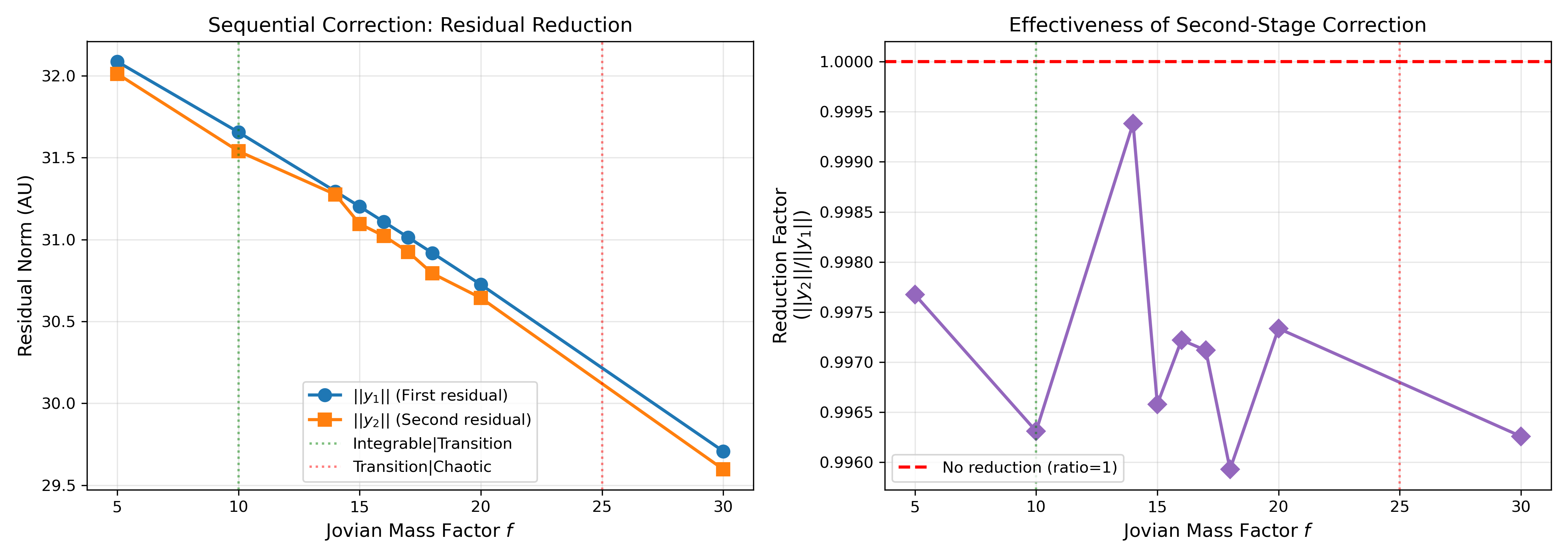}
\caption{Sequential correction analysis reveals negligible hierarchical decomposition. (Left) First-stage ($\|\mathbf{y}_1\|$, blue circles) and second-stage ($\|\mathbf{y}_2\|$, orange squares) residual norms. The two curves are nearly identical. (Right) Reduction ratio $\|\mathbf{y}_2\| / \|\mathbf{y}_1\|$. Values cluster near unity across all regimes, confirming that the reverse capacity transition reflects genuine simplification of the residual correction in the chaotic regime, not insufficient expressivity.}
\label{fig:sequential_corrections}
\end{figure*}

Figure~\ref{fig:sequential_corrections} shows sequential correction results. The refinement ratio $\|\mathbf{y}_2\| / \|\mathbf{y}_1\|$ remains close to unity across all regimes: integrable (mean 0.997), transitional (0.997), and chaotic (0.996). Even the large 3$\times$64 first-stage network leaves virtually no residual for the second-stage network to learn, confirming that single-stage networks already capture all learnable structure. The remaining residual $\mathbf{y}_2$ is dominated by irreducible stochasticity: trajectory-specific fluctuations arising from sensitive dependence on initial conditions that cannot be systematically compressed. This is the defining signature of ergodic smoothing---the learnable degrees of freedom of $\mathbf{y}(t)$ contract in the chaotic regime, leaving an irreducible noisy remainder that sequential refinement cannot reduce.

\subsection{Connection to Classical Chaos Theory}

KAM theory predicts that invariant tori survive perturbations below a critical threshold and disintegrate above it~\cite{laskar1989,murray1999}. Chirikov's resonance-overlap criterion~\cite{chirikov1979} predicts global chaos onset at $f \sim 10$--16 for our system. Our capacity transition aligns with this criterion both in location ($\hat{f}_c = 16.6 \pm 2.8$) and in character: the gradual onset with seed-dependent variability mirrors the physical picture of successive resonance overlaps rather than a sharp bifurcation.

\begin{figure*}[tb]
\centering
\includegraphics[width=0.95\textwidth]{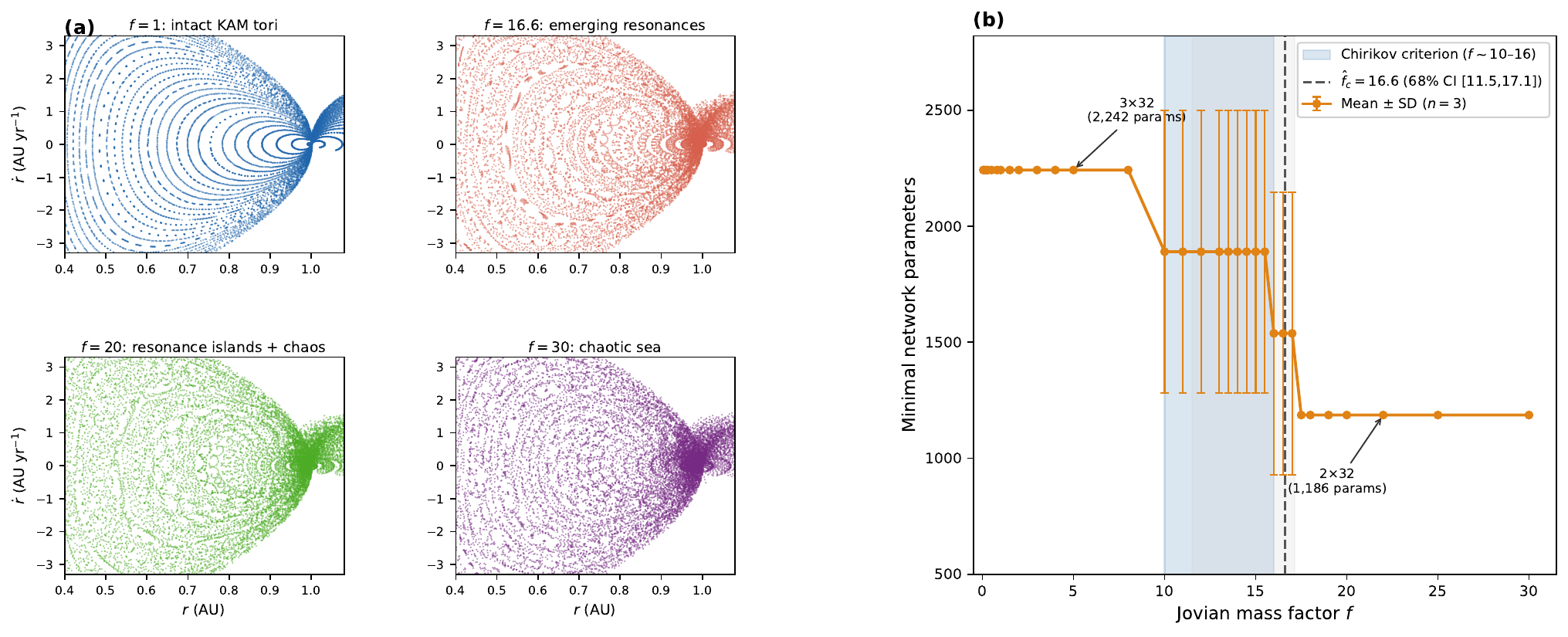}
\caption{Neural network capacity transitions align with incipient chaos onset as predicted by KAM theory. (a)~Poincar\'{e} sections at four representative mass factors. (b)~Capacity profile (orange line) exhibits non-monotonic behavior with peak at $f = 5$ and reverse drop at $f \geq 17$. The transition point (red band, $\hat{f}_c = 16.6 \pm 2.8$) aligns with Chirikov's criterion (blue band, $f \sim 10$--16), occurring at or before the onset of visible geometric chaos. The reverse capacity drop coincides with emergence of fully developed chaotic seas, consistent with ergodic smoothing.}
\label{fig:poincare_kam}
\end{figure*}

Figure~\ref{fig:poincare_kam} contextualizes our findings through Poincar\'{e} sections at four representative mass factors. At $f = 1$, the section shows a clean closed curve---an intact KAM torus. At $f = 16.6$, the structure remains largely regular with subtle distortions, consistent with incipient resonance overlap. At $f = 20$, prominent 3:1 resonance islands appear; at $f = 30$, widespread chaotic seas dominate.

Our capacity transition occurs at or before the onset of macroscopic geometric chaos. At $\hat{f}_c = 16.6$, some tori are beginning to distort under resonance overlap while others remain intact---precisely the regime where the residual $\mathbf{y}(t)$ must represent the most complex multi-scale structure. The reverse capacity drop at $f > 17$ highlights a fundamental decoupling between dynamical and representational complexity: classical dynamical complexity (Lyapunov exponents, Kolmogorov--Sinai entropy) increases monotonically with $f$, while representational complexity peaks at intermediate disorder and then \emph{decreases}. The alignment of $\hat{f}_c$ with Chirikov's criterion confirms that this decoupling is rooted in the physics of KAM torus destruction.

\section{Discussion}

\subsection{Perturbative Learning as a General Principle}

The 140-fold parameter reduction compared to full-trajectory prediction~\cite{breen2020} reflects a simple insight: networks should learn only the residual correction, not the complete signal. In the three-body problem, our networks learn only Jupiter's influence---the physically interesting component---achieving comparable accuracy with ${\sim}1{,}200$ parameters in both the integrable and fully chaotic regimes.

This principle extends far beyond celestial mechanics. In quantum chemistry, correcting Hartree-Fock or DFT baselines~\cite{McGibbon2017}; in fluid dynamics, learning turbulent corrections to laminar flows; in condensed matter physics, adding many-body corrections to mean-field band structures---all share this structure. Hybrid force-field/machine learning models demonstrate that augmenting analytical approximants with data-driven corrections can fully capture complex quantum-mechanical interactions and path-integral corrections~\cite{Fain1,Fain2}.

\subsection{Reverse Capacity Transition and Non-Monotonic Complexity}

Our central finding challenges conventional wisdom that chaos always increases learning difficulty. At the heart of this result is a distinction between two independently measurable notions of complexity. \emph{Dynamical complexity}---quantified by Lyapunov exponents, Kolmogorov--Sinai entropy, or fractal phase-space dimension---measures the asymptotic unpredictability of individual trajectories. \emph{Representational complexity}, as operationalized by our equalized-accuracy protocol, measures the minimal network capacity required to capture the learnable structure of $\mathbf{y}(t)$ to fixed accuracy. These two measures decouple at the chaos transition.

The global capacity maximum at $f = 5$ reflects a crossover between two competing effects: \emph{perturbation amplitude} and \emph{ergodic smoothing}. For $f \ll 5$, amplitude is too small to demand high capacity. For $f \gg 5$, ergodic smoothing reduces learnable content. The peak at $f = 5$ marks the crossover: large amplitude combined with strictly quasi-periodic structure, with no ergodic averaging to simplify the target.

In the transitional regime ($f \sim 15$--17), the residual $\mathbf{y}(t)$ must simultaneously encode (i) low-frequency quasi-periodic oscillations from surviving KAM tori and (ii) higher-frequency chaotic fluctuations from resonance overlap. This multi-scale superposition is the source of peak representational difficulty.

In the fully chaotic regime ($f \geq 17$), we term the resulting simplification \emph{ergodic smoothing}:
\begin{equation}
\mathbf{y}(t) = \bar{\mathbf{y}}(t) + \boldsymbol{\xi}(t),
\label{eq:decomp}
\end{equation}
where $\bar{\mathbf{y}}(t)$ is the statistically smooth ensemble-averaged correction a compact network can learn, and $\boldsymbol{\xi}(t)$ captures irreducible trajectory-specific fluctuations. The network learns only $\bar{\mathbf{y}}(t)$; $\boldsymbol{\xi}(t)$ encodes microscopic initial-condition information beyond the training distribution and cannot be systematically compressed. The sequential correction null result ($\|\mathbf{y}_2\|/\|\mathbf{y}_1\| \approx 0.997$) confirms this: $\mathbf{y}_2$ corresponds to the irreducible $\boldsymbol{\xi}(t)$ component of Eq.~\eqref{eq:decomp}.

\subsection{Implications for Neural Surrogate Design}

\textbf{(1) Intermediate-complexity regimes are most challenging.} Systems poised between order and chaos---turbulent boundary layers, weakly ionized plasmas, driven nonlinear oscillators---warrant careful capacity allocation, combining competing dynamical timescales that resist compact representation.

\textbf{(2) Chaos does not always require large capacity.} Ergodic smoothing implies that the learnable content of the residual saturates once the system is fully chaotic. This suggests opportunities for efficient surrogates in fully developed turbulence, mixing flows, and ergodic many-body systems.

\textbf{(3) Sequential correction provides negligible benefit.} Our null result validates the NNPT one-shot correction framework: single-stage networks capture all learnable structure, and hierarchical decomposition adds complexity without benefit.

\textbf{(4) Neural network capacity as a novel complexity diagnostic.} The equalized-accuracy protocol provides a computationally accessible proxy for representational complexity, distinct from classical chaos indicators. Minimal network capacity can be estimated from short training runs, suggesting a practical workflow: use capacity probing as a rapid diagnostic before committing to expensive full simulations.

Several promising directions emerge. Physics-informed architectures~\cite{raissi2019,karniadakis2021} that encode conservation laws or symmetries may navigate capacity barriers more efficiently. Graph neural networks~\cite{sanchez2020,battaglia2016} that respect interaction structure could scale better to many-body systems. Hybrid methods~\cite{ulibarrena2023} that dynamically switch between neural predictions and traditional solvers represent another avenue.

\subsection{Limitations and Future Directions}

\textbf{Generality across systems:} Does the reverse capacity transition appear in other chaotic systems? Preliminary investigations of driven pendulums and kicked rotors suggest similar non-monotonic patterns, but systematic studies across diverse physical domains are needed.

\textbf{Architecture dependence:} The capacity measurements reported here are specific to MLP architectures with SiLU activations. Whether alternative network classes exhibit the same non-monotonic transition remains an important open question. In particular, Kolmogorov-Arnold Networks (KANs)~\cite{liu2024kan}---which exploit compositional univariate function structure---may be better matched to the scalar-input geometry of the NNPT correction-learning problem. Applying the equalized-accuracy protocol to KAN architectures would provide an architecture-independent test of whether the reverse capacity transition is a robust physical phenomenon.

\textbf{Scaling to higher dimensions:} The three-body problem is low-dimensional ($d = 4$ for planar motion). Many-body systems and PDEs involve high-dimensional state spaces where capacity scaling may differ qualitatively.

\textbf{Connection to information theory:} The ergodic smoothing picture suggests a connection to rate-distortion theory: the irreducible $\boldsymbol{\xi}(t)$ component may correspond to information that cannot be compressed below a certain rate, while $\bar{\mathbf{y}}(t)$ lies in a low-dimensional manifold amenable to compact neural representation. Formalizing this connection~\cite{jacot2018,grassberger1986} remains an important open problem.

\textbf{Integration timescale and orbital confinement:} The 20-year integration window is short relative to the timescales on which KAM tori fully disintegrate and orbital escape occurs~\cite{laskar1989}. Over this window, Earth's positional displacement remains small (${\sim}0.03$~AU even at $f = 30$), so the orbit remains confined and chaos manifests primarily as functional complexity in $\mathbf{y}(t)$ rather than orbital instability. This is a deliberate feature of our experimental design. Whether the capacity transition persists or shifts at longer integration times is a natural direction for future work.

\section{Conclusion}

Using the three-body problem as an illustrative testbed, we established key principles for neural network surrogates of complex physical systems.

\textbf{First}, \emph{perturbative correction learning provides a general strategy for parameter-efficient surrogates}. By training networks to predict residuals after analytically subtracting tractable components, we achieve over 140-fold parameter reduction compared to full-trajectory prediction~\cite{breen2020}.

\textbf{Second}, \emph{physical complexity exhibits rich non-monotonic structure}. Our equalized-accuracy protocol reveals a capacity profile peaking in the late integrable regime ($f = 5$, 3$\times$32 with 2,242 parameters), remaining elevated through the transitional regime ($f \sim 15$--17), then \emph{decreasing} in the fully chaotic regime ($f \geq 17$, 2$\times$32 with 1,186 parameters)---a 47\% reduction from peak. The transition point $\hat{f}_c = 16.6 \pm 2.8$ aligns with Chirikov's resonance-overlap criterion~\cite{chirikov1979,lecar2001}. The physical mechanism is \emph{ergodic smoothing}: in the fully chaotic regime, trajectory-specific fluctuations become irreducible, leaving only a compact learnable residual.

\textbf{Third}, \emph{sequential correction provides negligible refinement}. The ratio $\|\mathbf{y}_2\| / \|\mathbf{y}_1\| \approx 0.997$ confirms that single-stage networks capture all learnable structure, validating the NNPT one-shot correction framework.

\textbf{Fourth}, \emph{neural network capacity and classical dynamical complexity are distinct, decoupled measures}. Dynamical complexity increases monotonically with disorder; representational complexity peaks at intermediate disorder and then decreases. This decoupling reveals that network capacity is a genuinely new complexity metric---sensitive to learnability structure rather than asymptotic unpredictability.

By introducing NNPT, we demonstrate that integrating exact solutions into the learning objective transforms networks from general function approximators into specialized perturbative operators. The reverse capacity transition reveals that chaos, far from universally increasing learning difficulty, exhibits rich non-monotonic structure that computational methods must navigate strategically.

\begin{acknowledgments}
We thank J.~P.~Sethna for affording us, long ago, with his pedagogical insights into KAM theory. Computations were performed using resources at Washington University in St.~Louis and Google Colaboratory. B.F. thanks Freecurve Labs for support. Z.N. acknowledges partial support from the Washington University SPEED grant program. Part of this work was performed at the Aspen Center for Physics, which is supported by National Science Foundation grant PHY-2210452. We are grateful to discussions with and comments by Denys Kononenko, Igor Kurnikov, Michael Levitt, Shmuel Nussinov, Leonid Pereyaslavets, and members of the Freecurve Labs team.
\end{acknowledgments}

\section*{Data and Code Availability}

Trajectory data, trained network weights, and analysis scripts are available at [REPOSITORY URL TO BE ADDED] and can be obtained from the corresponding author upon request.

\bibliographystyle{apsrev4-2}
\bibliography{main}

@book{BenderOrzag,
  author    = {Carl M. Bender and Steven A. Orszag},
  title     = {Advanced Mathematical Methods for Scientists and Engineers: Asymptotic Methods and Perturbation Theory},
  publisher = {Springer-Verlag},
  year      = {1999}
}

@book{KC,
  author    = {J. Kevorkian and J. D. Cole},
  title     = {Perturbation Methods in Applied Mathematics},
  series    = {Applied Mathematical Sciences},
  publisher = {Springer-Verlag},
  year      = {1981}
}

@book{Holmes,
  author    = {Mark H. Holmes},
  title     = {Introduction to Perturbation Methods},
  publisher = {Springer},
  year      = {2012},
  edition   = {2nd}
}

@book{Fernandez,
  author    = {Francisco M. Fernandez},
  title     = {Introduction to Perturbation Theory in Quantum Mechanics},
  publisher = {CRC Press},
  year      = {2020}
}

@book{LL,
  author    = {L. D. Landau and E. M. Lifshitz},
  title     = {Quantum Mechanics, Non-relativistic Theory},
  series    = {Course of Theoretical Physics},
  volume    = {3},
  edition   = {2nd},
  publisher = {Pergamon Press},
  year      = {1965}
}

@article{UAP1,
  author  = {G. Cybenko},
  title   = {Approximation by superpositions of a sigmoidal function},
  journal = {Mathematics of Control, Signals, and Systems},
  volume  = {2},
  pages   = {303--314},
  year    = {1989}
}

@article{UAP2,
  author  = {K. Hornik and M. Stinchcombe and H. White},
  title   = {Multilayer feedforward networks are universal approximators},
  journal = {Neural Networks},
  volume  = {2},
  pages   = {359--366},
  year    = {1989}
}

@book{QFT,
  author    = {Michel Talagrand},
  title     = {What Is a Quantum Field Theory?},
  publisher = {Cambridge University Press},
  year      = {2022}
}

@article{Fain1,
  author  = {A. Illarionov and S. Sakipov and L. Pereyaslavets and I. V. Kurnikov and G. Kamath and O. Butin and E. Voronina and I. Ivahnenko and I. Leontyev and G. Nawrocki and M. Darkhovskiy and M. Olevanov and Y. K. Cherniavskyi and C. Lock and S. Greenslade and S. K. Sankaranarayanan and M. G. Kurnikova and J. Potoff and R. D. Kornberg and M. Levitt and B. Fain},
  title   = {Accurate Force Field for Molecular Dynamics Simulation by Machine Learning},
  journal = {Journal of the American Chemical Society},
  volume  = {145},
  number  = {41},
  pages   = {23620--23638},
  year    = {2023},
  doi     = {10.1021/jacs.3c08562}
}

@article{Fain2,
  author  = {I. V. Kurnikov and L. Pereyaslavets and G. Kamath and S. N. Sakipov and E. Voronina and O. Butin and A. Illarionov and I. Leontyev and G. Nawrocki and M. Darkhovskiy and M. Olevanov and I. Ivahnenko and Y. Chen and C. B. Lock and M. Levitt and R. D. Kornberg and B. Fain},
  title   = {ARROW Force Field for Protein Simulations},
  journal = {Journal of Chemical Theory and Computation},
  volume  = {20},
  number  = {3},
  pages   = {1347--1359},
  year    = {2024},
  doi     = {10.1021/acs.jctc.3c01051}
}

@article{breen2020,
  author  = {P. G. Breen and C. N. Foley and T. Boekholt and S. Portegies Zwart},
  title   = {Newton versus the machine: solving the chaotic three-body problem using deep neural networks},
  journal = {Monthly Notices of the Royal Astronomical Society},
  volume  = {494},
  number  = {2},
  pages   = {2465--2470},
  year    = {2020},
  doi     = {10.1093/mnras/staa713}
}

@article{greydanus2019,
  author  = {Samuel Greydanus and Misko Dzamba and Jason Yosinski},
  title   = {Hamiltonian Neural Networks},
  journal = {Advances in Neural Information Processing Systems},
  volume  = {32},
  pages   = {15379--15389},
  year    = {2019}
}

@inproceedings{cranmer2020,
  author    = {Miles Cranmer and Sam Greydanus and Stephan Hoyer and Peter Battaglia and David Spergel and Shirley Ho},
  title     = {Lagrangian Neural Networks},
  booktitle = {ICLR 2020 Workshop on Integration of Deep Neural Models and Differential Equations},
  year      = {2020}
}

@article{poincare1890,
  author  = {Henri Poincar{\'e}},
  title   = {Sur le probl{\`e}me des trois corps et les {\'e}quations de la dynamique},
  journal = {Acta Mathematica},
  volume  = {13},
  pages   = {1--270},
  year    = {1890},
  doi     = {10.1007/BF02392506}
}

@article{laskar1989,
  author  = {Jacques Laskar},
  title   = {A numerical experiment on the chaotic behaviour of the Solar System},
  journal = {Nature},
  volume  = {338},
  pages   = {237--238},
  year    = {1989},
  doi     = {10.1038/338237a0}
}

@book{murray1999,
  author    = {Carl D. Murray and Stanley F. Dermott},
  title     = {Solar System Dynamics},
  publisher = {Cambridge University Press},
  address   = {Cambridge},
  year      = {1999}
}

@book{sethna2006,
  author    = {James P. Sethna},
  title     = {Entropy, Order Parameters, and Complexity},
  publisher = {Oxford University Press},
  address   = {Oxford},
  year      = {2006}
}

@inproceedings{sanchez2020,
  author    = {Alvaro Sanchez-Gonzalez and Jonathan Godwin and Tobias Pfaff and Rex Ying and Jure Leskovec and Peter Battaglia},
  title     = {Learning to Simulate Complex Physics with Graph Networks},
  booktitle = {Proceedings of the 37th International Conference on Machine Learning},
  series    = {ICML 2020},
  year      = {2020}
}

@article{schutt2017,
  author  = {K. T. Sch{\"u}tt and F. Arbabzadah and S. Chmiela and K. R. M{\"u}ller and A. Tkatchenko},
  title   = {Quantum-chemical insights from deep tensor neural networks},
  journal = {Nature Communications},
  volume  = {8},
  pages   = {13890},
  year    = {2017},
  doi     = {10.1038/ncomms13890}
}

@article{reichstein2019,
  author  = {Markus Reichstein and Gustau Camps-Valls and Bjorn Stevens and Martin Jung and Joachim Denzler and Nuno Carvalhais and Prabhat},
  title   = {Deep learning and process understanding for data-driven Earth system science},
  journal = {Nature},
  volume  = {566},
  pages   = {195--204},
  year    = {2019},
  doi     = {10.1038/s41586-019-0912-1}
}

@article{jin2020,
  author  = {Pengzhan Jin and Zhen Zhang and Aiqing Zhu and Yifa Tang and George Em Karniadakis},
  title   = {{SympNets}: Intrinsic structure-preserving symplectic networks for identifying Hamiltonian systems},
  journal = {Neural Networks},
  volume  = {132},
  pages   = {166--179},
  year    = {2020},
  doi     = {10.1016/j.neunet.2020.08.017}
}

@article{Batzner2022,
  author  = {Simon Batzner and Albert Musaelian and Lixin Sun and Mario Geiger and Jonathan P. Mailoa and Mordechai Kornbluth and Nicola Molinari and Tess E. Smidt and Boris Kozinsky},
  title   = {{E(3)}-equivariant graph neural networks for data-efficient and accurate interatomic potentials},
  journal = {Nature Communications},
  volume  = {13},
  pages   = {2453},
  year    = {2022},
  doi     = {10.1038/s41467-022-29939-5}
}

@article{Kondor2025,
  author  = {Risi Kondor},
  title   = {Equivariant Neural Networks and Representation Theory},
  journal = {Proceedings of the National Academy of Sciences},
  volume  = {122},
  year    = {2025},
  note    = {In press}
}

@article{raissi2019,
  author  = {Maziar Raissi and Paris Perdikaris and George Em Karniadakis},
  title   = {Physics-informed neural networks: A deep learning framework for solving forward and inverse problems involving nonlinear partial differential equations},
  journal = {Journal of Computational Physics},
  volume  = {378},
  pages   = {686--707},
  year    = {2019},
  doi     = {10.1016/j.jcp.2018.10.045}
}

@article{karniadakis2021,
  author  = {George Em Karniadakis and Ioannis G. Kevrekidis and Lu Lu and Paris Perdikaris and Sifan Wang and Liu Yang},
  title   = {Physics-informed machine learning},
  journal = {Nature Reviews Physics},
  volume  = {3},
  pages   = {422--440},
  year    = {2021},
  doi     = {10.1038/s42254-021-00314-5}
}

@article{chirikov1979,
  author  = {Boris V. Chirikov},
  title   = {A universal instability of many-dimensional oscillator systems},
  journal = {Physics Reports},
  volume  = {52},
  number  = {5},
  pages   = {263--379},
  year    = {1979},
  doi     = {10.1016/0370-1573(79)90023-1}
}

@book{hairer2006,
  author    = {Ernst Hairer and Christian Lubich and Gerhard Wanner},
  title     = {Geometric Numerical Integration: Structure-Preserving Algorithms for Ordinary Differential Equations},
  edition   = {2nd},
  publisher = {Springer},
  address   = {Berlin},
  year      = {2006},
  doi       = {10.1007/3-540-30666-8}
}

@article{elfwing2018,
  author  = {Stefan Elfwing and Eiji Uchibe and Kenji Doya},
  title   = {Sigmoid-weighted linear units for neural network function approximation in reinforcement learning},
  journal = {Neural Networks},
  volume  = {107},
  pages   = {3--11},
  year    = {2018},
  doi     = {10.1016/j.neunet.2017.12.012}
}

@inproceedings{kingma2015,
  author    = {Diederik P. Kingma and Jimmy Ba},
  title     = {Adam: A Method for Stochastic Optimization},
  booktitle = {Proceedings of the 3rd International Conference on Learning Representations},
  series    = {ICLR 2015},
  year      = {2015},
  note      = {arXiv:1412.6980}
}

@article{McGibbon2017,
  author  = {Robert T. McGibbon and Andrew G. Taube and Alexander G. Donchev and Karthik Siva and Felipe Hern{\'a}ndez and Cory Hargus and Karl Law and John L. Klepeis and David E. Shaw},
  title   = {Improving the accuracy of M{\o}ller-Plesset perturbation theory with neural networks},
  journal = {The Journal of Chemical Physics},
  volume  = {147},
  number  = {16},
  pages   = {161725},
  year    = {2017},
  doi     = {10.1063/1.4986081}
}

@inproceedings{battaglia2016,
  author    = {Peter W. Battaglia and Razvan Pascanu and Matthew Lai and Danilo Jimenez Rezende and Koray Kavukcuoglu},
  title     = {Interaction Networks for Learning about Objects, Relations and Physics},
  booktitle = {Advances in Neural Information Processing Systems},
  volume    = {29},
  pages     = {4502--4510},
  year      = {2016}
}

@article{ulibarrena2023,
  author  = {V. Saz Ulibarrena and P. Horn and S. Portegies Zwart and E. Sellentin and B. Koren and M. X. Cai},
  title   = {Hybrid integration of {$N$}-body simulations with deep neural networks},
  journal = {Journal of Computational Physics},
  volume  = {496},
  pages   = {112565},
  year    = {2024},
  doi     = {10.1016/j.jcp.2023.112565}
}

@article{lecar2001,
  author  = {Myron Lecar and Fred A. Franklin and Matthew J. Holman and Norman W. Murray},
  title   = {Chaos in the Solar System},
  journal = {Annual Review of Astronomy and Astrophysics},
  volume  = {39},
  pages   = {581--631},
  year    = {2001},
  doi     = {10.1146/annurev.astro.39.1.581}
}

@article{Liao2022,
  author  = {Shijun Liao and Xiaoming Li and Yu Yang},
  title   = {Three-body problem -- from Newton to supercomputer plus machine learning},
  journal = {New Astronomy},
  volume  = {96},
  pages   = {101850},
  year    = {2022}
}

@book{cover2006,
  author    = {Thomas M. Cover and Joy A. Thomas},
  title     = {Elements of Information Theory},
  edition   = {2nd},
  publisher = {Wiley-Interscience},
  address   = {Hoboken, NJ},
  year      = {2006},
  doi       = {10.1002/047174882X}
}

@inproceedings{jacot2018,
  author    = {Arthur Jacot and Franck Gabriel and Cl{\'e}ment Hongler},
  title     = {Neural Tangent Kernel: Convergence and Generalization in Neural Networks},
  booktitle = {Advances in Neural Information Processing Systems},
  volume    = {31},
  pages     = {8571--8580},
  year      = {2018}
}

@article{grassberger1986,
  author  = {Peter Grassberger},
  title   = {Toward a quantitative theory of self-generated complexity},
  journal = {International Journal of Theoretical Physics},
  volume  = {25},
  number  = {9},
  pages   = {907--938},
  year    = {1986},
  doi     = {10.1007/BF00668821}
}

@article{liu2024kan,
  title={{KAN}: {K}olmogorov-{A}rnold Networks},
  author={Liu, Ziming and Wang, Yixuan and Vaidya, Sachin and 
          Ruehle, Fabian and Halverson, James and Solja{\v{c}}i{\'c}, 
          Marin and Hou, Thomas Y and Tegmark, Max},
  journal={arXiv preprint arXiv:2404.19756},
  year={2024}
}
\nocite{*}
\end{document}